\documentstyle[11pt]{article}
\begin{document}

\title{\textbf{Resonantly damped oscillations of elliptically shaped stratified emerging coronal loops}}
\author{K. Karami$^{1}$\thanks{KKarami@uok.ac.ir} , S. Amiri${^1}\thanks{Sirwanamiri@yahoo.co.uk}$ ,
K. Bahari${^2}\thanks{K.Bahari@razi.ac.ir}$ , Z.
Ebrahimi${^1}\thanks{zanyar.ebrahimi@gmail.com}$\\$^{1}$\small{Department
of Physics, University of Kurdistan, Pasdaran Street, Sanandaj,
Iran}\\$^{2}$\small{Physics Department, Faculty of Science, Razi
University, Kermanshah, Iran}}

\maketitle

\begin{abstract}
The effects of both elliptical shape and stage of emergence of the
coronal loop on the resonant absorption of standing kink
oscillations are studied. To do so, a typical coronal loop is
modeled as a zero-beta longitudinally stratified cylindrical
magnetic flux tube. We developed the connection formulae for the
resonant absorption of standing transversal oscillations of a
coronal loop with an elliptical shape, at various stages of its
emergence. Using the connection formulae, the dispersion relation is
derived and solved numerically to obtain the frequencies and damping
rates of the fundamental and first-overtone kink modes. Our
numerical results show that both the elliptical shape and stage of
emergence of the loop alter the frequencies and damping rates of the
tube as well as the ratio of frequencies of the fundamental and its
first-overtone modes. However, the ratio of the oscillation
frequency to the damping rate is not affected by the tube shape and
stage of its emergence and also is independent of the density
stratification parameter.
\end{abstract}


\noindent{\textit{Key words:} Sun: corona --- Sun: magnetic fields
--- Sun: oscillations}
\clearpage
\section{Introduction}

Solar corona and its extraordinary high temperature has been the
topic of various debates and studies from several decades ago. The
origin and the source of coronal continual heating and high
temperature have been related to coronal loops. The claim that the
coronal loops and their behaviors such as their damping oscillations
may be one of the main reasons of coronal heating, has been
investigated through several studies so far.

Transverse oscillations of coronal loops have been observed by the
Transition Region and Coronal Explorer (TRACE) for several years
(see e.g. Aschwanden et al. 1999; Schrijver $\&$ Brown 2000).
Nakariakov et al. (1999) interpreted these oscillations as fast kink
modes with the period ranging from $2.3$ to $10.8$ min and decay
time from $3.2$ to $20.8$ min. The observed values of the periods
and decay times make it possible to obtain indirect information on
the conditions of the plasma and magnetic field in coronal loops.

Ofman $\&$ Aschwanden (2002) used the data deduced by Aschwanden et
al. (2002) to investigate the oscillations of 11 coronal loops. They
argued that the observed TRACE loops consist of multiple unresolved
thin loop threads which produce inhomogeneous internal structure of
the observed loop. They adopted 1-dimensional cartesian slabs of
plasma with the magnetic field lines in the $z$-direction and the
direction of the inhomogeneity along the $x$-axis normal to the
magnetic surfaces, as a simple model for the oscillating loops. They
found that the dependence of the decay time on both the length $L$
and the width $w$ of the loop is in excellent agreement with the
power law damping predicted by phase mixing.

 The property of resonant absorption as a non-thermal mechanism
 makes it possible to describe the heating of magnetic loops in solar corona as well as rapid
 decaying of magnetohydrodynamics (MHD) waves even in weakly dissipative plasmas
 (see e.g. Ionson 1978; Poedts et al. 1989; Ofman et al. 1994; Erd\'{e}lyi \& Goossens 1994, 1995; Tirry $\&$
 Goossens 1996; Andries et al. 2005b; Safari et al. 2006; Dymova $\&$ Ruderman 2006; Goossens et al. 2009).

Verwichte et al. (2004), using the observations of TRACE, have
identified the fundamental and its first harmonic of the transverse
kink mode in two coronal loops. The period ratios observed by
Verwichte et al. (2004) are $1.81 \pm 0.25$ and $1.64 \pm 0.23$.
However, these values were corrected with the improvement of the
observational error bars to $1.82 \pm 0.08$ and $1.58 \pm 0.06$,
respectively, by Van Doorsselaere et al. (2007). Also Verth et al.
(2008) added some further corrections by considering the effects of
loop expansion and estimated a period ratio of 1.54. All these
values clearly are lower than 2. This may be caused by different
factors such as the effects of density stratification (see e.g.
Andries et al. 2005a; Erd\'{e}lyi \& Verth 2007; Karami \& Asvar
2007; Safari et al. 2007; Karami et al. 2009) and magnetic twist
(see Erd\'{e}lyi \& Carter 2006; Erd\'{e}lyi \& Fedun 2006; Karami
\& Barin 2009; Karami \& Bahari 2010, 2012) in the loops.

Karami et al. (2009, hereafter Paper I) investigated the effect of
longitudinally stratification on resonant absorption of MHD waves
for both kink $(m=1)$ and fluting $(m=2)$ modes. They found that the
frequencies and damping rates of both the fundamental and
first-overtone modes increase when the stratification parameter
increases. Also for stratified loops they obtained the ratio of the
frequencies $\omega_2/\omega_1$ of the first overtone and its
fundamental mode less than 2.

Morton \& Erd\'{e}lyi (2009) studied the effects of both the
elliptical shape and stage of emergence of the loops on the period
ratio $P_1/P_2$ for the minor and major elliptical cases. Their
results showed that the parameter characterising the stage of
emergence does affect the value of period ratio $P_1/P_2$.
Particularly, the greatest contribution from emergence to the period
ratio occurs when the loop is fully emerged. Also they showed that
the ellipticity of the loop has an important role in the value of
$P_1/P_2$ for minor elliptical case but the major ellipse was found
to have a less effect on the period ratio of standing
oscillations.

Here we combine the two models considered in Paper I and Morton $\&$
Erd\'{e}lyi (2009) to investigate of the effects of both elliptical
shape and stage of emergence of the coronal loop on the resonant
absorption of standing transversal kink oscillations observed by the
TRACE. This paper is organized as follows. In Sections 2 and 3, we
combine the two techniques of Paper I and Morton $\&$ Erd\'{e}lyi
(2009) to derive the equations of motion, introduce the relevant
connection formulae and obtain the dispersion relation. In Section
4, we give numerical results. Section 5 is devoted to conclusions.

\section{Equations of motion and modeling of the flux tube}

The linearized MHD equations for a zero-beta plasma are given by
\begin{eqnarray}
\frac{\partial\delta{\mathbf v}}{\partial
t}=\frac{1}{4\pi\rho}\{(\nabla\times\delta{\mathbf B})\times{\mathbf
B} +(\nabla\times{\mathbf B})\times\delta{\mathbf B}\}
+\frac{\eta}{\rho}\nabla^2\delta{\mathbf v},\label{mhd1}
\end{eqnarray}
\begin{eqnarray}
\frac{\partial\delta{\mathbf B}}{\partial
t}=\nabla\times(\delta{\mathbf v}\times{\mathbf B})+
\frac{c^2}{4\pi\sigma}\nabla^2\delta{\mathbf B},\label{mhd2}
\end{eqnarray}
where $\delta\bf{v}$ and $\delta\bf{B}$ are the Eulerian
perturbations of velocity and magnetic fields; $\bf{B}$, $\rho$,
$\sigma$, $\eta$ and $c$ are the background magnetic filed, the mass
density, the electrical conductivity, the viscosity and the speed of
light, respectively.

The simplifying assumptions are the same as in Karami $\&$ Asvar
(2007). According to Andries et al. (2005b) and Paper I, one can
expand the perturbed quantities $\delta\bf{v}$ and
 $\delta\bf{B}$ as follows
\begin{eqnarray}
\delta{\mathbf B}(r,z)=\sum_{k=1}^{\infty}\delta{\mathbf B}^{(k)}(r)\psi^{(k)}(z),\nonumber\\
\delta{\mathbf v}(r,z)=\sum_{k=1}^{\infty}\delta{\mathbf
v}^{(k)}(r)\psi^{(k)}(z),~~\label{Brz}
\end{eqnarray}
where $\psi^{(k)}(z)$s form a complete set of orthonormal
 eigenfunctions and satisfy the eigenvalue relation
 \begin{eqnarray}
 L_{A}\psi^{(k)}=\zeta_{k}\psi^{(k)},\label{alf}
 \end{eqnarray}
where $L_{A}$ is the Alfv\'{e}n operator,
\begin{eqnarray}
L_A=\rho\omega^2+\frac{B^2}{4\pi}\frac{\partial^2}{\partial
z^2}=\rho\left(\omega^2+v_A^2\frac{\partial^2}{\partial z^2}\right),
\end{eqnarray}
with Alfv\'{e}n velocity $v_A=\frac{B}{\sqrt{4\pi\rho}}$ and
straight constant background magnetic filed ${\bf B}=B\hat{{\bf
z}}$.

We further assume there is a density stratification along the tube
axis in $z$-direction.Since we are interested in resonantly
 damped oscillations, it implies that the density profile must be radially structured
  too. Following Paper I, we consider the density profile given by
\begin{eqnarray}
\rho(r,z)=\rho_0(r)\rho(z),
\end{eqnarray}
where
\begin{eqnarray}
\rho_0(r)=\left\{\begin{array}{ccc}
    \rho_{{\rm in}},&(r<R_1),&\\
    \Big[\frac{\rho_{\rm in}-\rho_{\rm ex}}{R-R_1}\Big](R-r)+\rho_{\rm ex},&(R_1<r< R),&\\
      \rho_{{\rm ex}},&(r>R).&\\
      \end{array}\right.
\end{eqnarray}
Here, $R$ is the loop radius and $R_1<R$ is the radius of the
homogeneous part of the tube. The radius at which resonant
absorption occurs is between $R_1$ and $R$. The thickness of the
inhomogeneous layer, $l=R -R_1$, will be assumed to be small. Here,
$\rho_{\rm in}$ and $\rho_{\rm ex}$ are the footpoint densities of
the interior and exterior regions of the tube, respectively.

According to Morton $\&$ Erd\'{e}lyi (2009) we consider two types of
elliptical loop that can occur, the minor elliptical loop where
minor axis of the ellipse is the vertical axis of the loop, and the
major elliptical loop where major axis of the ellipse is the
vertical axis of the loop. Note that the minor ellipse is a
situation that occurs most plausibly under coronal conditions.

For the minor elliptical case, the longitudinally stratified density
 profile takes the form
\begin{eqnarray}
\rho\big(z\big)=\exp\left(-\mu\frac{\cos\Big(\alpha(z)\Big)
\left[1-\epsilon^{2}\sin^{2}\Big(\alpha(z)\Big)\right]^{-\frac{1}{2}}-\lambda}{1-\lambda}\right),\label{rhozminor}
\end{eqnarray}
and for the major one, it is given by
\begin{eqnarray}
\rho\big(z\big)=\exp\left(-\mu\frac{\cos\Big(\alpha(z)\Big)\Big(1-\epsilon^2\Big)^{1/2}
\left[1-\epsilon^{2}\cos^{2}\Big(\alpha(z)\Big)\right]^{-\frac{1}{2}}-\lambda}{1-\lambda}\right),\label{rhozmajor}
\end{eqnarray}
where
\begin{eqnarray}
\epsilon=\left(1-\frac{b^2}{a^2}\right)^{1/2},
\end{eqnarray}
is the ellipticity of the loop with minor half-axis of length $b$,
and major half-axis of length $a$. Also $\mu:=\frac{L}{\pi H}$ is
defined as stratification parameter, where $H$ and $L$ are the
density scale height and length of the loop, respectively. The
parameter $\lambda$ describes the stage of emergence of the loop
from the photosphere. It is defined as the ratio of the distance of
the photosphere from center of the ellipse to the vertical
half-axis. A positive value of $\lambda$ refers to the situation in
which the center of ellipse is sitting below the photosphere (early
stage emergence), and thus, the negative $\lambda$ for the center
above the photosphere (late stage emergence). A zero value of
$\lambda$ corresponds to a loop having a semi-elliptical shape. For
the minor elliptical case, $\lambda$ is given by
\begin{eqnarray}
\lambda=1-\frac{\mu H}{b},
\end{eqnarray}
where $\mu H$ is the distance of the loop apex from the photosphere.
For the major one, $\lambda$ is defined as
\begin{eqnarray}
\lambda=1-\frac{\mu H}{a}.
\end{eqnarray}
Note we have considered a tube of length $L$ which its footpoints
are
  in the two points $z=0$ and $z=L$, and also note that in Eqs. (\ref{rhozminor}) and (\ref{rhozmajor}),
$\alpha(z)$ is the angle between the vertical axis of the loop and
the line joining the center of the ellipse to the plasma element
located at distance $z$ along the tube. Following Morton $\&$
Erd\'{e}lyi (2009) for the minor elliptical case, one can obtain the
value of $\alpha(z)$ by calculating the ellipse arc length defined
as
\begin{eqnarray}
\int^{t_1}_{0}\Big(1-\epsilon^2\sin^2(t)\Big)^{\frac{1}{2}}{\rm
d}t=\left(\frac{2z}{L}-1\right)
\int^{t_2}_{0}\Big(1-\epsilon^2\sin^2(t)\Big)^{\frac{1}{2}}{\rm
d}t,\label{integ1}
\end{eqnarray}
where $t_1$ and $t_2$ are parametric angles given by
\begin{eqnarray}
t_1=\arctan\left(\frac{b}{a}\tan(\alpha)\right),~~~t_2=\arctan\left(\frac{b}{a}\tan(\theta)\right),\label{t1}
\end{eqnarray}
and
\begin{equation}
\theta=\arctan\left[\frac{1}{\lambda}\left(\frac{1-\lambda^2}{1-\epsilon^2}\right)^{\frac{1}{2}}\right],\label{Bzrin}
\end{equation}
is the angle between the vertical axis of the loop and a line that
joins the ellipse center to the loop foot-point (see Fig. 2 in
Morton $\&$ Erd\'{e}lyi 2009).

For the major elliptical case, we have
\begin{eqnarray}
\int^{\frac{\pi}{2}}_{t_1}\Big(1-\epsilon^2\sin^2(t)\Big)^{\frac{1}{2}}{\rm
d}t=\left(\frac{2z}{L}-1\right)\int^{\frac{\pi}{2}}_{t_2}
\Big(1-\epsilon^2\sin^2(t)\Big)^{\frac{1}{2}}{\rm d}t,\label{integ2}
\end{eqnarray}
where
\begin{eqnarray}
t_1=\arctan\left(\frac{b}{a}\cot(\alpha)\right),~~~t_2=\arctan\left(\frac{b}{a}\cot(\theta)\right),\label{t1maj}
\end{eqnarray}
and
\begin{equation}
\theta=\arctan\left[\frac{\Big((1-\epsilon^2)(1-\lambda^2)\Big)^{\frac{1}{2}}}{\lambda}\right].\label{Bzrin}
\end{equation}
According to Paper I, in the absence of dissipation, in the interior
region $(r<R_1)$, solutions of Eqs. (\ref{mhd1}) and (\ref{mhd2})
are
\begin{eqnarray} \delta
B_z^{(\rm{in})}(r,z)=\sum_{k=1}^{+\infty}A^{\rm{(in,k)}}J_{\rm
m}(|k_{\rm{in,k}}|r)\psi^{(\rm{in,k})}(z),\label{soli1}
\end{eqnarray}
\begin{eqnarray}
\delta v_r^{(\rm{in})}(r,z)=-\frac{i\omega
B}{4\pi}\sum_{k=1}^{+\infty}\frac{k_{\rm{in,k}}}{\zeta_{\rm{in,k}}}A^{(\rm{in,k})}J'_{\rm
m}(|k_{\rm{in,k}}|r)\psi^{(\rm{in,k})}(z),\label{soli}
\end{eqnarray}
where
\begin{eqnarray}
k_{\rm{in,k}}^2=\frac{\zeta_{\rm{ in,k} }}{B^2/4\pi}.
\end{eqnarray}
Here $J_{\rm {m}}$ is the Bessel function of the first kind and a
prime on $J_{\rm{m}}$ and hereafter on each function indicates a
derivative with respect to their appropriate arguments. The
solutions for the exterior region $r>R$, are the same as equation
(\ref{soli}) except that $J_{\rm m}$, index $\rm{"in"}$, and $|k_{
\rm{in,k}}|$ are replaced by $K_{\rm m}$, $\rm{"ex"}$ and $k_{\rm{
ex,k}}=-\frac{\zeta_{\rm{ex,k} }}{B^2/4\pi}$, respectively,
everywhere. Where $K_m$ is the modified Bessel function of the
second kind and shows that the wave amplitude vanishes in large
distance away from the tube boundary.

\section{Boundary conditions, Connection formulae and dispersion relation }

In the absence of dissipation effects, an appropriate dispersion
relation is obtained by requiring that the solutions for perturbed
quantities are continues at the tube surface. When a dissipative
layer is considered, the solutions may experience jumps across the
layer. An appropriate relation connecting the solutions of outside
and inside the tube, is called the ``connection formulae''.
Following Paper I, the jump across the boundary (resonance layer)
for $\delta B_z$ and $\delta v_r$ is
\begin{eqnarray}
\left[\delta B_z\right]= 0,\label{jumps1}
\end{eqnarray}
\begin{eqnarray}
 \left[\delta v_r
 \right]=-\sum^{+\infty}_{k=1}\frac{B\tilde{\omega} m^2\left\langle {{\phi^{\rm_{(in,k)}}}}
 \mathrel{\left| {\vphantom {{\phi^{\rm_{(in,k)}}} {\delta B_z^{\rm_{(in,k)}}}}}
 \right. \kern-\nulldelimiterspace}
 {{\delta B_z^{\rm{(in,k)}}}} \right\rangle}{4r_A^2\left\langle {{\phi^{\rm{(in,k)}}}}
 \mathrel{\left | {\vphantom {{\phi^k} {\L_{A1}\phi^{\rm{(in,k)}}}}}
 \right. \kern-\nulldelimiterspace}
 {{L_{A1}\Big |\phi^{\rm{(in,k)}}}} \right\rangle}\phi^{\rm{(in,k)}},\label{jumps}
 \end{eqnarray}
 where
\begin{eqnarray}
 L_{A1}=\frac{\partial L_A}{\partial r}\Big|_{r=r_A}=\tilde{\omega}^2[1+S_{kk}]\frac{\partial\rho_0(r)}{\partial
r}\Big|_{r=r_A},\label{La1}
\end{eqnarray}
and
\begin{eqnarray}
\phi^{\rm{(in,k)}}=\sqrt{\frac{2}{L}}\sum^{+\infty}_{j=1}\phi^{\rm{(in,k)}}_j\sin\left(\frac{j\pi}{L}z\right),
\end{eqnarray}
with
\begin{eqnarray}
\phi _{j}^{\rm{(in,k)}}  = \left\{ {\begin{array}{*{20}c}
   {\frac{{k^2 S_{kj} }}{{\rho_{\rm{in}}(1 + S_{kk} )(j^2  - k^2 )}}} & {j \ne k}  \\
   1 & {j = k}  \\
\end{array}}.\right.\label{phi}
\end{eqnarray}
Here $\phi^{\rm_{(in,k)}}$ satisfies $L_{A}\phi^{\rm_{(in,k)}}=0$
and $\tilde{\omega}=\omega+i\gamma$ where $\gamma$ is damping rate.
Also $R_1<r_A<R$ $R_1<r_{\rm A}<R$ is the radius at which the
singularity occurs. Note that $l=R-R_1$ is the thickness of the
inhomogeneous layer and Davila (1987) showed that in the resonance
absorption, however, the damping rate is independent of the
dissipation coefficient values. But the resonance layer width scales
as $\delta_{\rm
A}\propto(\frac{\nu}{\rho}+\frac{c^2}{4\pi\sigma})^{1/3}$. Karami \&
Bahari (2010) showed that for the Reynolds
${\mathcal{R}}=\Big(\frac{{\rm R}^2\rho_{\rm
i}}{\nu}\Big)/\Big(\frac{2\pi {\rm R}}{v_{A_{\rm i}}}\Big)=560$ and
Lundquist $S=\Big(\frac{4\pi\sigma {\rm
R}^2}{c^2}\Big)/\Big(\frac{2\pi {\rm R} }{v_{A_{\rm i}}}\Big)=10^4$
numbers given by Ofman et al. (1994), and taking $L=10^5$ km,
$R/L=0.01$, $a/R=0.08$, $\rho_{\rm ex}/\rho_{\rm in}=0.1$, and
interior Alfv\'{e}n velocity $v_{A_{\rm in}}=2000$ km s$^{-1}$ for a
typical coronal loop, then one can get $\delta_{\rm A}\simeq 85$ km
which is very close to the thickness of the inhomogeneous layer
$l\simeq 87$ km. This suggests that one can use the thin boundary
approximation which assumes that the thickness of the resonance
layer is the same as the inhomogeneous layer width (see also
Goossens et al. 2009).

Note that the effects of elliptical shape and stage of emergence of
the loop on the resonant absorption appear in jump conditions via
the function $S_{kj}$ which is obtained as
\begin{eqnarray}
S_{kj}=\sqrt{\frac{2}{L}}\int_0^L\sin\left(\frac{k\pi}{L}
z\right)\ln{\Big(\rho(z)\Big)}\sin\left(\frac{j\pi}{L} z\right){\rm
d}z .
 \end{eqnarray}
Substituting the fields of equation (\ref{soli1}) and (\ref{soli})
in jump conditions (\ref{jumps1}) and (\ref{jumps}) gives
\begin{eqnarray}
 \left( {\begin{array}{*{20}c}
   {\Pi_1^{(\rm ex,1)} } & { - \Pi_1^{(\rm in,1)} } & {\Pi_1^{(\rm ex,2)} } & { - \Pi_1^{(\rm in,2)} } &  \ldots   \\
   {\Xi_1^{(\rm ex,1)} } & {\Xi_1^{(\rm in,1)}  + {\mathcal{D}}_1^{(\rm in,1)} } & {\Xi_1^{(\rm ex,2)} } & {\Xi_1^{(\rm in,2)}  + {\mathcal{D}}_1^{(\rm in,2)} } &  \ldots   \\
   {\Pi_2^{(\rm ex,1)} } & { - \Pi_2^{(\rm in,1)} } & {\Pi_2^{(\rm ex,2)} } & { - \Pi_2^{(\rm in,2)} } &  \ldots   \\
   {\Xi_2^{(\rm ex,1)} } & {\Xi_2^{(\rm in,1)}  + {\mathcal{D}}_2^{(\rm in,1)} } & {\Xi_2^{(\rm ex,2)} } & {\Xi_2^{(\rm in,2)}  + {\mathcal{D}}_2^{(\rm in,2)} } &  \ldots   \\
    \vdots  &  \vdots  &  \vdots  &  \vdots  &  \ddots   \\
\end{array}}\right)\left( {\begin{array}{*{20}c}
   {A^{(\rm ex,1)} }  \\
   {A^{(\rm in,1)} }  \\
   {A^{(\rm ex,2)} }  \\
   {A^{(\rm in,2)} }  \\
    \vdots   \\
\end{array}} \right)
= 0,\label{disperssion}
\end{eqnarray}
where all the definitions in Eq. (\ref{disperssion}) are as those of
Paper I. Note that the dispersion relation is obtained by requiring
that the system (\ref{disperssion}) has non-trivial solutions, i.e.
its determinant is zero. In the next section, we solve the
dispersion relation numerically to obtain the frequencies $\omega$
and damping rates $\gamma$ of the resonantly damped MHD kink
oscillations of longitudinally stratified elliptical emerging
coronal loops.
\section{Numerical Results}

To solve the dispersion relation (\ref{disperssion}) numerically, we
chose the physical parameters $L=10^5~{\rm km}$, $R/L=0.01$,
$l/R=0.02$, $B=100~{\rm G}$, $\rho_{\rm{in}}=2\times10^{-14}~{\rm
g~cm^{-3}}$ and $\rho_{\rm{ex}}/\rho_{\rm{in}}=0.1$. For such a loop
one finds $v_{A_{\rm{in}}}=\frac{B}{\sqrt{4\pi\rho_{\rm{in}}}}=2\times 10^3~{\rm km~s^{-1}}$ and
$\omega_{A_{\rm{in}}}:= \frac{v_{A_{\rm{in}}}}{L}=0.02~{\rm
rad~s^{-1}}$. In what follows, we illustrate our numerical studies
in the three separate equilibrium cases containing (i) circle-arc
emerged loop (ii) minor elliptical semi-emerged loop and (iii) minor
and major elliptical loops.

\subsection{Circle-arc emerged loop ($\epsilon=0$, $\lambda\neq0$)}

\begin{figure}
\includegraphics{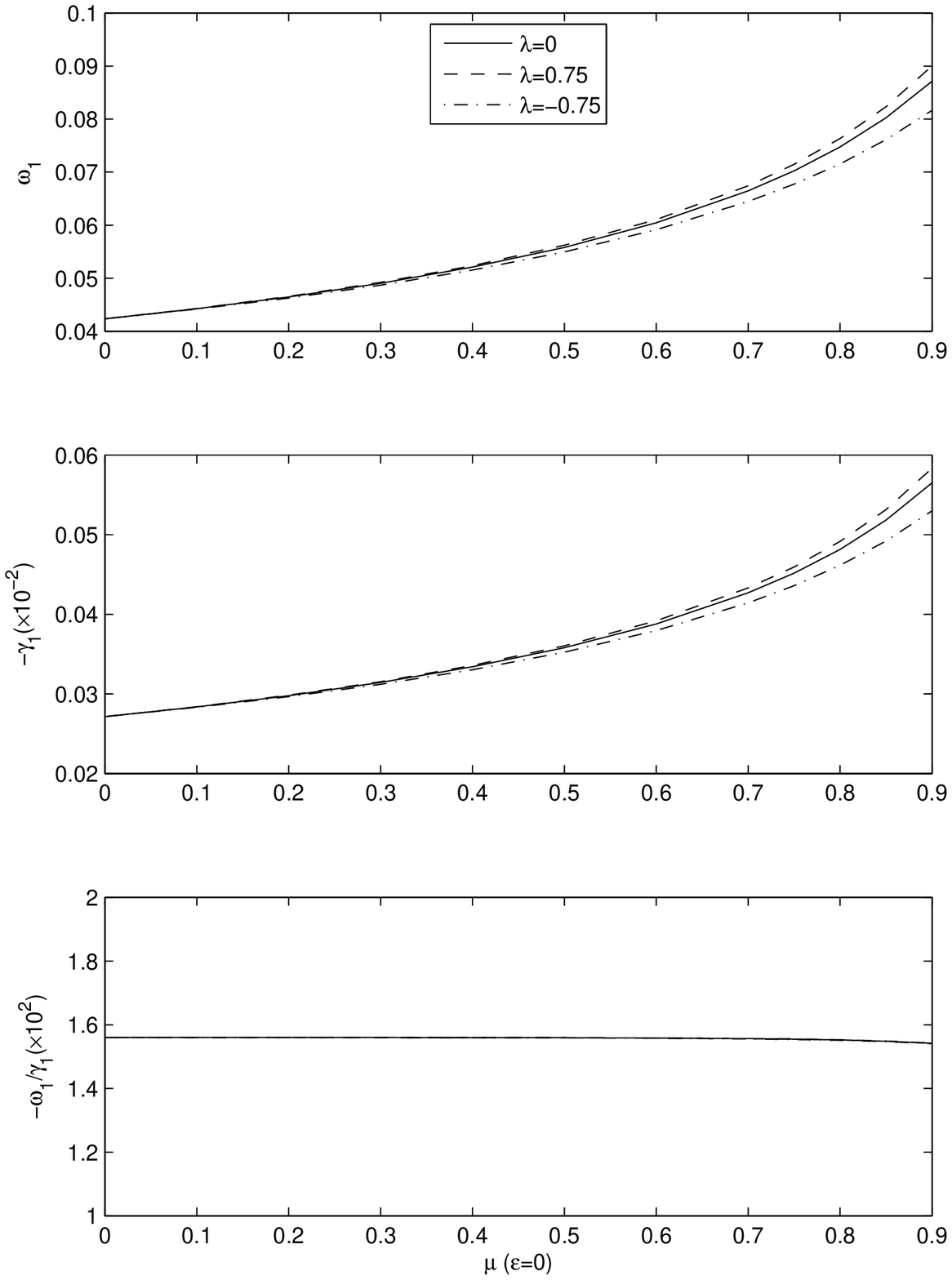}
 \vspace{5.5cm}
\caption[] {Frequency of the fundamental kink ($m=1$) mode and its
damping rate as well as the ratio of the oscillation frequency to
the damping rate as a function of the stratification parameter $\mu$
for a circle-arc flux tube ($\epsilon=0$) at various stages of
emergence $\lambda=0$ (solid), $0.75$ (dashed) and $-0.75$
(dash-dotted). The loop parameters are $L=10^5~\rm{km}$, $R/L=0.01$,
$l/R=0.02$, $\rho_{\rm{ex}}/\rho_{\rm{in}}=0.1$,
$\rho_{\rm{in}}=2\times10^{-14} \rm{gr~cm^{-3}}$ and $B=100~{\rm
G}$. Both frequencies and damping rates are in units of the interior
Alfv\'{e}n frequency, $\omega_{\rm A_{in}}=0.02{\rm~rad~s^{-1}}$.}
\label{V1}
 \end{figure}

\begin{figure}
\includegraphics{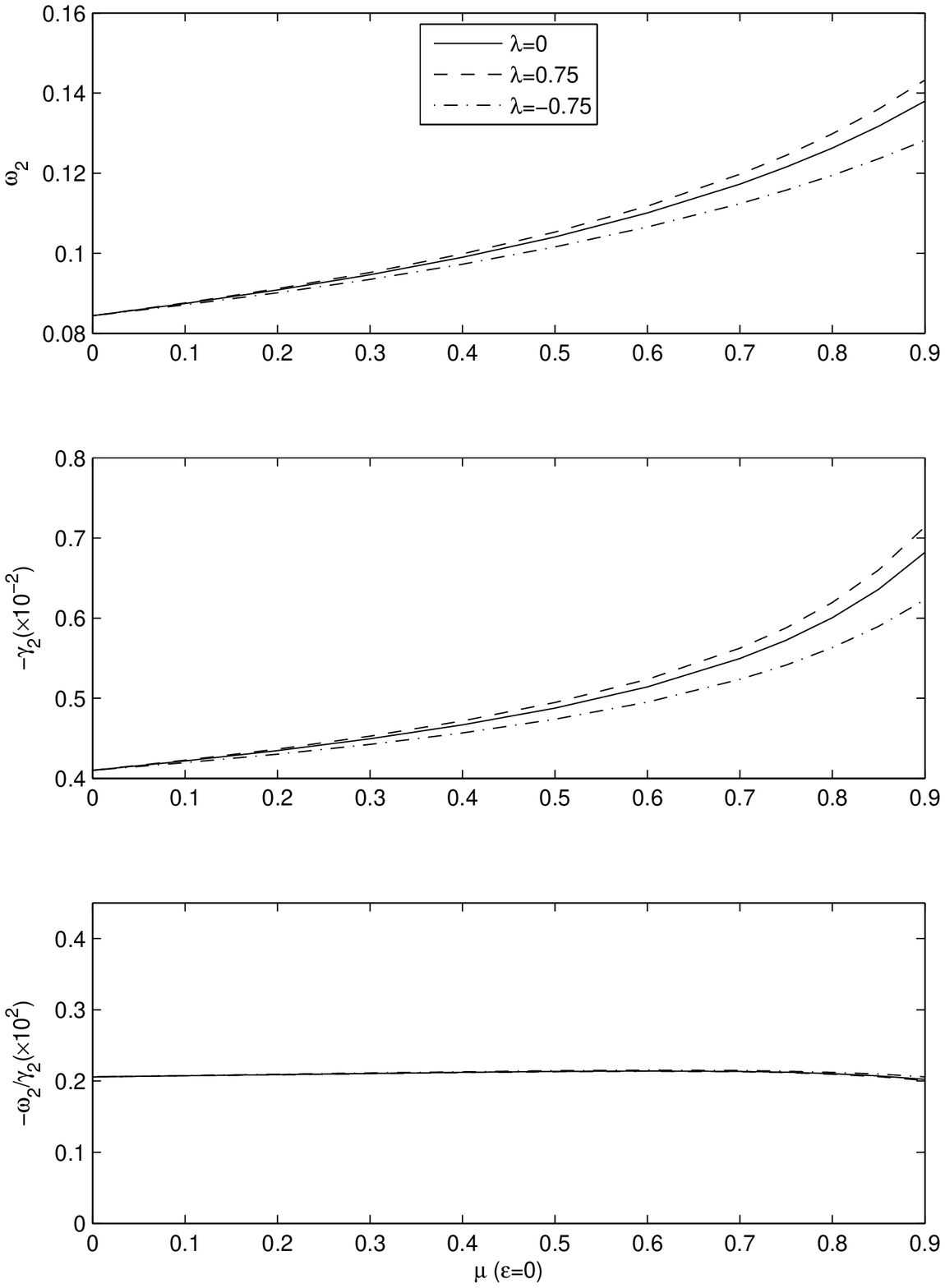}
 \vspace{5.5cm}
\caption[] {Same as Fig. \ref{V1}, for the first-overtone kink
      modes.}
\label{V2}
 \end{figure}

The effect of stage of emergence of the tube on both the frequencies
$\omega$ and damping rates $\gamma$ are calculated by numerical
solution of the dispersion relation, i.e. Eq. (\ref{disperssion}).
In Figs. \ref{V1} and \ref{V2}, the frequencies, damping rates and
their ratio for the fundamental and first-overtone kink $(m=1)$
modes are plotted versus the stratification parameter $\mu$ for a
circle-arc flux tube ($\epsilon=0$) at various stages of emergence
containing early stage emergence ($\lambda=0.75$), semi-emerged
ellipse ($\lambda=0$) and late stage emergence ($\lambda=-0.75$).
Figures \ref{V1} and \ref{V2} show that (i) for a given loop shape
parameter $\lambda$, both frequencies $\omega_1$, $\omega_2$ and
their corresponding damping rates $|\gamma_1|$, $|\gamma_2|$
increase when the stratification parameter $\mu$ increases. (ii) For
a given $\mu$, both frequencies and damping rates increase when
$\lambda$ increases. For instance, for $\mu=0.9$, the early stage
emergence $\lambda =0.75$ in comparison with late stage emergence
$\lambda =-0.75$ would cause $\omega_1, \omega_2, |\gamma_1|$ and
$|\gamma_2|$ to increase by about $10.2\%, 11.7\%, 10\%$ and
$14.5\%$, respectively. (iii) The ratio of the oscillation frequency
to the damping rate, $\omega/|\gamma|$, is independent of
stratification. Also the stage of emergence of the loop does not
affect $\omega/|\gamma|$. This means that the ratio
$\omega/|\gamma|$ is independent of how the loop center is situated
with respect to the photosphere.

\begin{figure}
\includegraphics{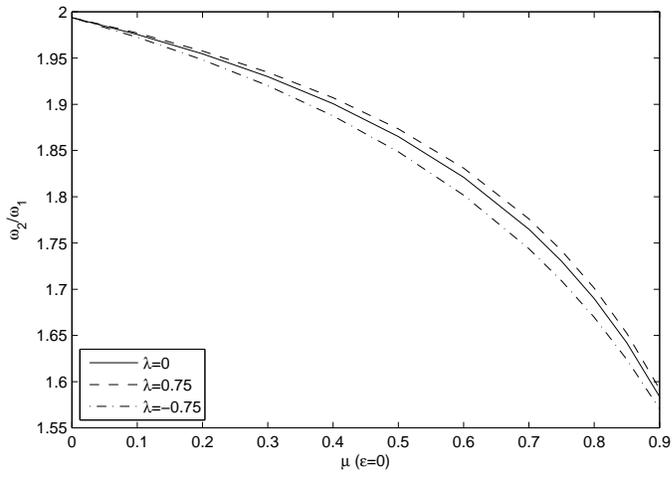}
 \vspace{5.5cm}
\caption[]{Ratio of the frequencies $\omega_2/\omega_1$ of the
first-overtone and its fundamental kink ($m=1$) mode versus
      $\mu$ for a circle-arc flux tube ($\epsilon=0$) at various
stages of emergence $\lambda=0$ (solid), $0.75$ (dashed) and $-0.75$
(dash-dotted). Auxiliary parameters as in Fig. \ref{V1}.} \label{V3}
 \end{figure}

In Fig. \ref{V3}, the ratio of the frequencies $\omega_2/\omega_1$
of the first overtone and its fundamental mode is plotted versus the
stratification parameter. Figure \ref{V3} shows that (i) for a given
$\lambda$, the frequency ratio decreases from 2 (for an unstratified
loop) with increasing density stratification. (ii) For a given
$\mu$, the frequency ratio decreases when $\lambda$ decreases. The
results of $\omega_2/\omega_1$ are in agreement with those obtained
by Morton $\&$ Erd\'{e}lyi (2009).

\subsection{Minor elliptical semi-emerged loop ($\epsilon\neq0$,
$\lambda=0$)}

\begin{figure}
\includegraphics{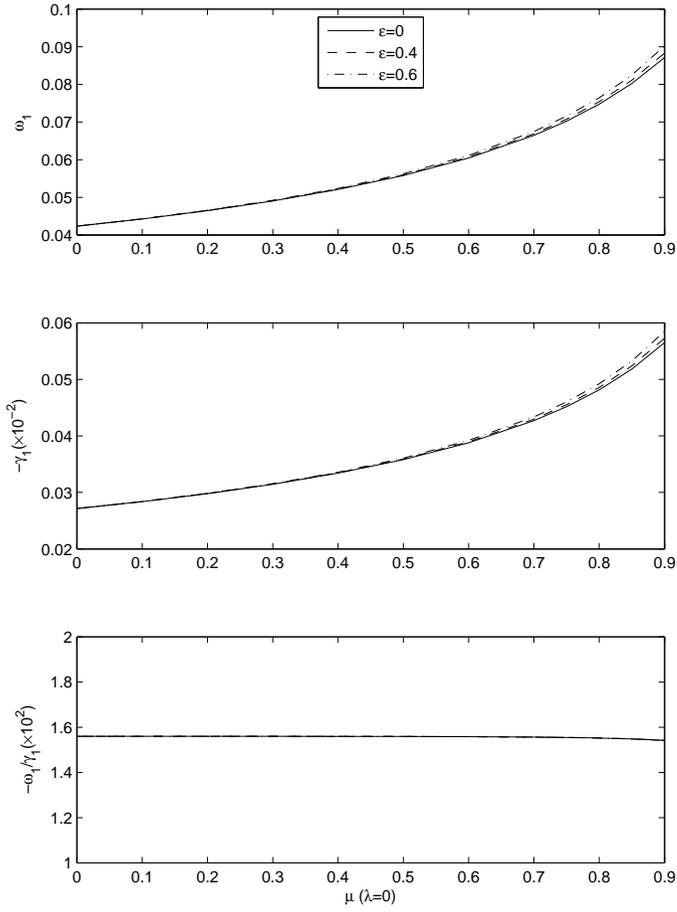}
 \vspace{5.5cm}
\caption[]{Frequency of the fundamental kink ($m=1$) mode and its
damping rate as well as the ratio of the oscillation frequency to
the damping rate as a function of the stratification parameter $\mu$
for a minor semi-emerged loop ($\lambda=0$) with different
ellipticity parameters $\epsilon=0.0$ (solid), $0.4$ (dashed) and
$0.6$ (dash-dotted). Auxiliary parameters as in Fig. \ref{V1}.}
\label{V4}
 \end{figure}

\begin{figure}
\includegraphics{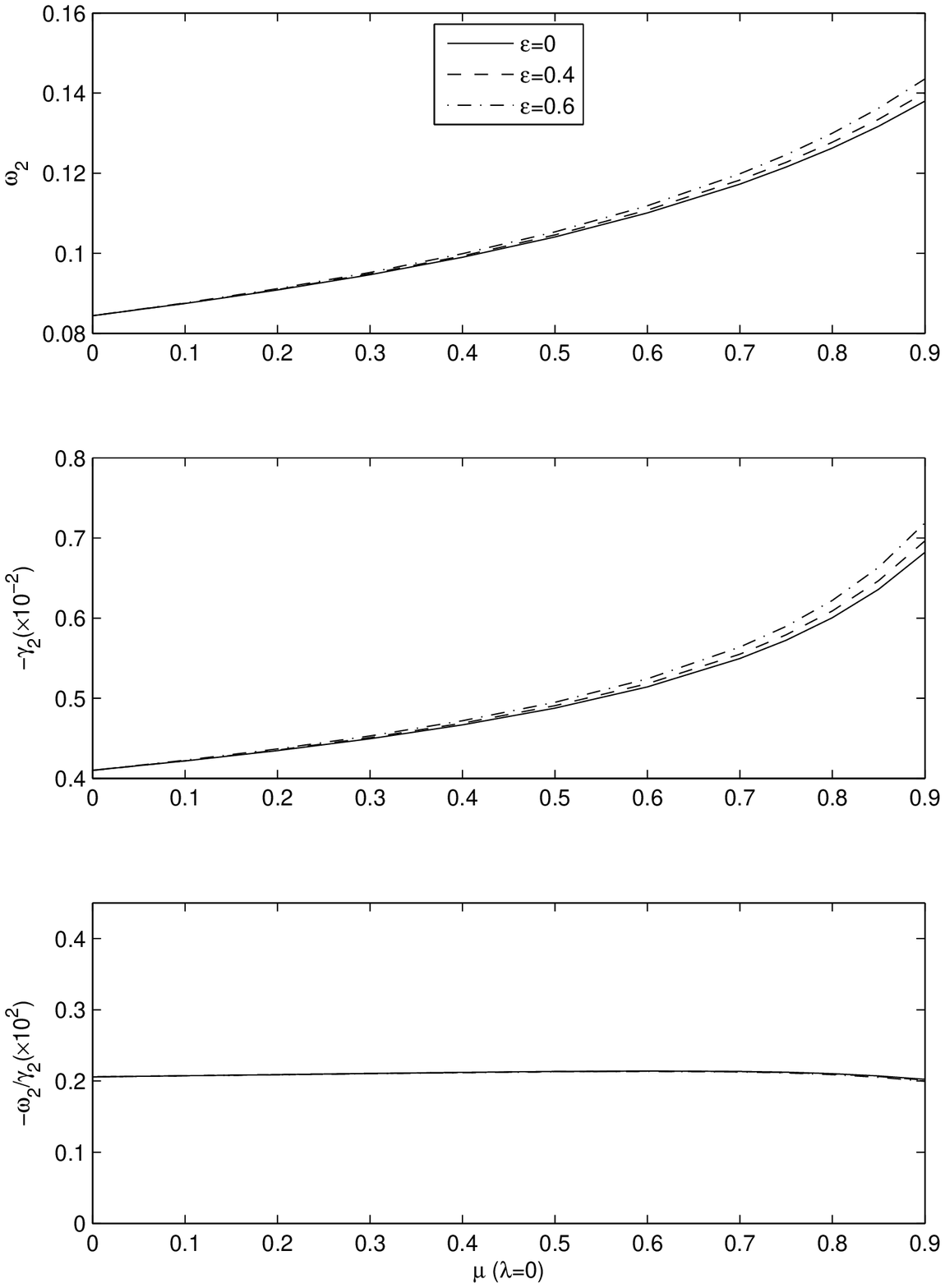}
 \vspace{5.5cm}
\caption[]{Same as Fig. \ref{V4}, for the first-overtone kink
      modes.}
\label{V5}
 \end{figure}

Figures \ref{V4} and \ref{V5} as Figs. \ref{V1} and \ref{V2}
display the result of frequency, damping rate and ratio
$\omega/|\gamma|$ but for a minor semi-emerged loop ($\lambda=0$)
with different ellipticity parameters $\epsilon=0.0$, $0.4$ and
$0.6$. Figures \ref{V4} and \ref{V5} show that (i) for a given
ellipticity parameter $\epsilon$, both frequencies $\omega_1$,
$\omega_2$ and their corresponding damping rates $|\gamma_1|$,
$|\gamma_2|$ increase when the stratification parameter $\mu$
increases. (ii) For a given $\mu$, both frequencies and damping
rates increase with increasing $\epsilon$. For instance, for $\mu =
0.9$ considering a loop with $\epsilon=0.6$ would cause to increase
$\omega_1$, $\omega_2$, $|\gamma_1|$ and $|\gamma_2|$ up to $3.5\%$,
$4\%$, $3.5\%$ and $5.3\%$, respectively, in comparison with a
semi-circular loop ($\epsilon=0$). (iii) The ratio $\omega/|\gamma|$
remains unchanged by increasing the stratification parameter. Also
this ratio is independent of the elliptical shape of the loop.

\begin{figure}
\includegraphics{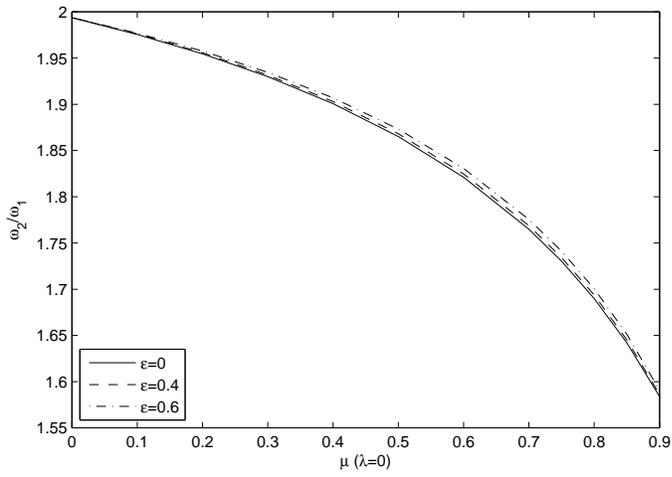}
 \vspace{5.5cm}
\caption[]{Ratio of the frequencies $\omega_2/\omega_1$ of the
first-overtone and its fundamental kink ($m=1$) mode versus
      $\mu$ for a minor semi-emerged loop ($\lambda=0$) with different
ellipticity parameters $\epsilon=0.0$ (solid), $0.4$ (dashed) and
$0.6$ (dash-dotted). Auxiliary parameters as in Fig. \ref{V1}.}
\label{V6}
 \end{figure}

Figure \ref{V6} illustrates the frequency ratio $\omega_2/\omega_1$
as a function of stratification parameter for a semi-emerged loop
($\lambda=0$) with different ellipticity parameters. Figure \ref{V6}
presents that (i) for a given ellipticity parameter, the ratio
$\omega_2/\omega_1$ decreases from 2 by increasing the
stratification parameter and approaches below 1.5. (ii) For a given
$\mu$, the frequency ratio $\omega_2/\omega_1$ decreases when
$\epsilon$ decreases. This is in good concord with the result of
Morton $\&$ Erd\'{e}lyi (2009).

\subsection{Minor and major elliptical emerged loops ($\epsilon\neq0$, $\lambda\neq0$)}

\begin{figure}
\includegraphics{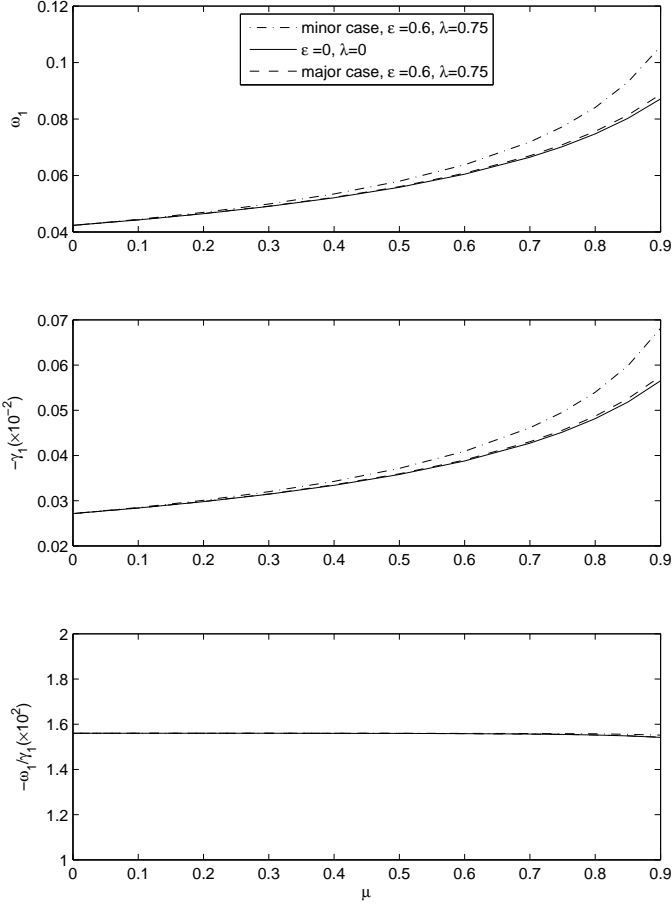}
 \vspace{5.5cm}
\caption[] {Frequency of the fundamental kink ($m=1$) mode and its
damping rate as well as the ratio of the oscillation frequency to
the damping rate as a function of the stratification parameter $\mu$
for circular-arc semi-emerged (solid), minor (dash-dotted) and major
(dashed) elliptical loops. Auxiliary parameters as in Fig.
\ref{V1}.} \label{V7}
 \end{figure}

\begin{figure}
\includegraphics{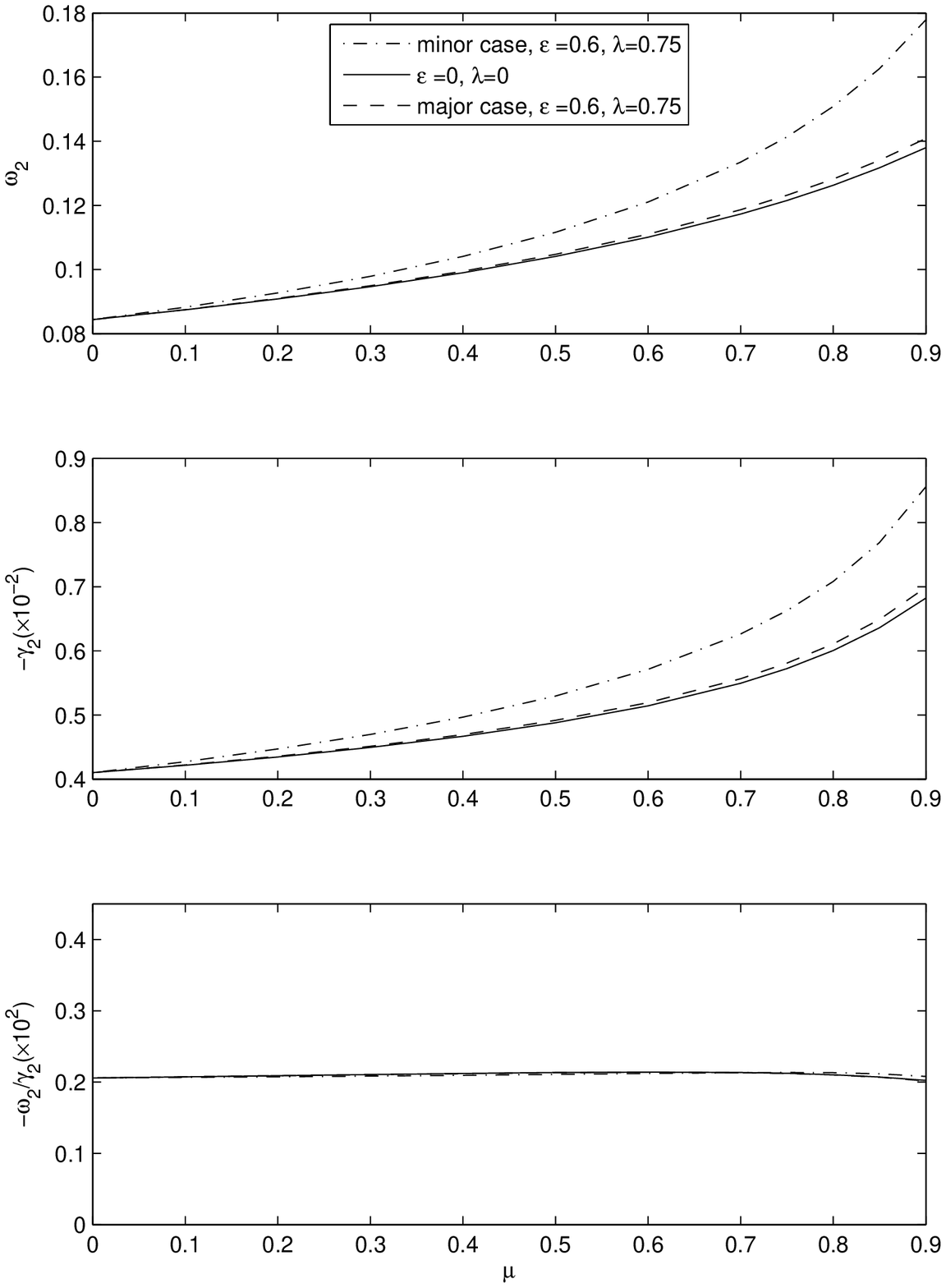}
 \vspace{5.5cm}
\caption[] {Same as Fig. \ref{V7}, for the first-overtone kink
      modes.}
\label{V8}
 \end{figure}

\begin{figure}
\includegraphics{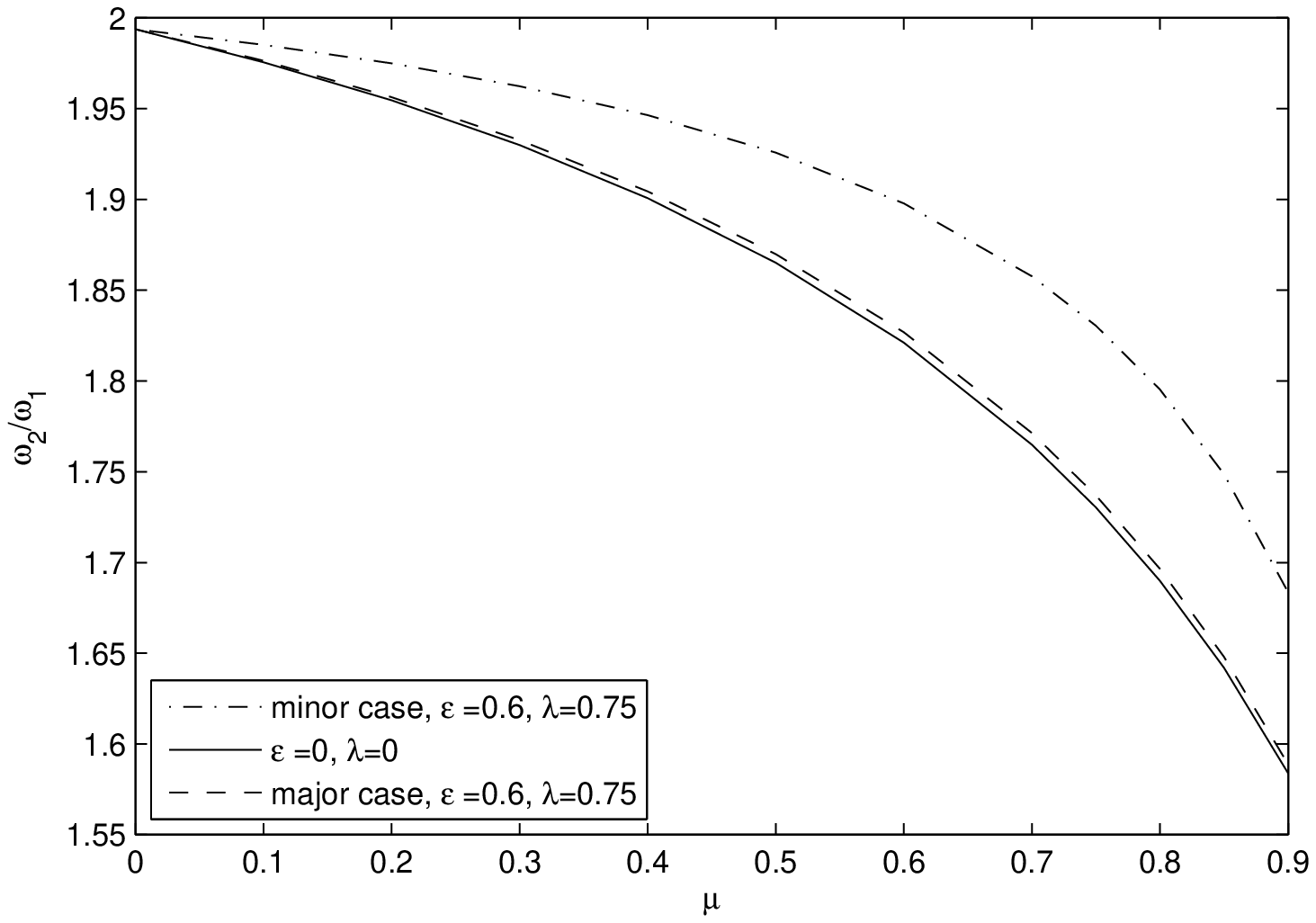}
 \vspace{5.5cm}
\caption[]{Ratio of the frequencies $\omega_2/\omega_1$ of the
first-overtone and its fundamental kink ($m=1$) mode versus
      $\mu$ for circular-arc semi-emerged (solid),
minor (dash-dotted) and major (dashed) elliptical loops. Auxiliary
parameters as in Fig. \ref{V1}.} \label{V9}
 \end{figure}

Figures \ref{V7}, \ref{V8} and \ref{V9} make it possible to do some
comparisons between major and minor elliptical cases. Figures
\ref{V7} and \ref{V8} show that: (i) the frequencies and damping
rates of both fundamental and first-overtone kink ($m=1$) modes of a
minor elliptical case are greater than those of a major elliptical
case and also of a circular-arc semi-emerged loop
($\epsilon=\lambda=0$) studied in Paper I. For instance, for
$\mu=0.9$ for a minor elliptical loop with $\epsilon = 0.6$ and
$\lambda = 0.75$, the values of $\omega_1$, $\omega_2$, $|\gamma_1|$
and $|\gamma_2|$ are $19.1\%$, $26.2\%$, $18.4\%$ and $22.4\%$
greater than those of a major elliptical case, respectively. (ii)
The frequencies and damping rates of a major elliptical loop are
slightly greater than the circular-arc semi-emerged loop. (iii) The
ratio $\omega/|\gamma|$ of both fundamental and first-overtone modes
do not affected by stratification for minor/major elliptical and
circular-arc semi-emerged loops.

In Fig. \ref{V9}, the frequency ratio $\omega_2/\omega_1$ versus the
stratification parameter is plotted for major and minor cases as
well as a circular-arc semi-emerged loop. Figure \ref{V9} shows
that: (i) for a given $\mu$, the frequency ratio $\omega_2/\omega_1$
of the minor elliptical loop is greater than the major one. For
instance, for $\mu=0.9$, the ratio $\omega_2/\omega_1$ of a minor
elliptical case is $5.9\%$ greater than that of a major elliptical
one. (ii) The ratio $\omega_2/\omega_1$ in a major elliptical loop
is slightly greater than the circular-arc semi-emerged loop.
Therefore, the major ellipse has a lesser effect on the frequency
ratio of standing kink oscillations. This is in good accordance with
that obtained by Morton $\&$ Erd\'{e}lyi (2009).

\section{Conclusions}

Here we investigated the effects of both ellipticity and stage of
emergence on the resonant absorption of standing kink waves in
longitudinally stratified coronal loops. We considered a coronal
loop as a pressureless cylindrical flux tube embedded in a straight
magnetic field that undergoes a longitudinal density stratification
and radial density structuring. We extended the relevant connection
formulae for the resonant absorption of transverse kink oscillations
of a coronal loop with an elliptical shape, at various stages of its
emergence from the sub-photosphere into the solar corona. We studied
three stages of emergence of the loop at which the center of
eclipse is sitting below ($\lambda>0$), on ($\lambda=0$), and above
($\lambda<0$) the photosphere. We considered two types of elliptical
loop that can occur, the minor ellipse and the major ellipse. Note
that the minor ellipse is a situation that occurs most plausibly
under coronal conditions. By numerically solving the dispersion
relation, we obtained the frequencies and damping rates of the
fundamental and first-overtone kink ($m=1$) modes. Our numerical
results show the following.

 i) By increasing the density stratification parameter
$\mu$ in the loop, both frequencies and damping rates increase while
the frequency ratio $\omega_2/\omega_1$ decreases.

ii) In a circle-arc emerged loop ($\epsilon=0$, $\lambda\neq0$) for
a given $\mu$, the frequencies, damping rates and the frequency
ratio increase when the stage of emergence parameter $\lambda$
increases.

iii) In a minor elliptical semi-emerged loop ($\epsilon\neq0$,
$\lambda=0$) for a given $\mu$, the frequencies, damping rates and
the frequency ratio increase when the ellipticity parameter
$\epsilon$ increases.

iv) In a minor elliptical emerged loop ($\epsilon\neq0$,
$\lambda\neq0$)
 for a given $\mu$, the frequencies, damping rates and
frequency ratio are greater than those of a major one. However, the
results obtained for the aforementioned quantities in the
 major elliptical emerged loop are slightly greater than those
of a circular-arc semi-emerged loop ($\epsilon=\lambda=0$).

v) The ratio of the oscillation frequency to the damping rate,
$\omega/|\gamma|$, in minor/major elliptical and circular-arc
semi-emerged loops, is not affected by making changes in density
stratification parameter, ellipticity and stage of emergence of the
loop.
\section*{Acknowledgements}
The work of K. Karami has been supported financially by Department
of Physics, University of Kurdistan, Sanandaj, Iran under research
project No. 1/1390.

\clearpage
\clearpage
\clearpage
\clearpage
\clearpage
\clearpage
\clearpage
\clearpage
\clearpage

\begin{thebibliography}{00}
\bibitem{andr2005a} Andries J., Arregui I., Goossens M., 2005a, ApJ, 624, L57

\bibitem{andr2005b} Andries J., Goossens M., Hollweg J.V., Arregui I., Van Doorsselaere T., 2005b, A$\&$A, 430, 1109

\bibitem{asch2002} Aschwanden M.J., De Pontieu B., Schrijver C.J., Title A.M., 2002, Sol. Phys., 206, 99

\bibitem{asch1999} Aschwanden M.J., Fletcher L., Schrijver C.J., Alexander D., 1999, ApJ, 520, 880


\bibitem{Davila} Davila J.M., 1987, ApJ, 317, 514

\bibitem{dym2006} Dymova M.V., Ruderman M.S., 2006, A$\&$A, 457, 1059


\bibitem{Erdelyi} Erd\'{e}lyi R., Goossens M., 1994, Ap\&SS, 213,
273

\bibitem{Erdelyi} Erd\'{e}lyi R., Goossens M., 1995, A\&A, 294, 575


\bibitem{Erdelyi2007} Erd\'{e}lyi R., Verth G., 2007, A\&A, 462,
743

\bibitem{Goossens} Goossens M., Terradas J., Andries J., Arregui I., Ballester J.L., 2009, A\&A, 503,
213

\bibitem{ion1978} Ionson J.A., 1978, ApJ, 226, 650

\bibitem{Karami} Karami K., Asvar A., 2007, MNRAS, 381, 97

\bibitem{Karami} Karami K., Barin M., 2009, MNRAS, 394, 521

\bibitem{Karami} Karami K., Nasiri S., Amiri S., 2009,
MNRAS, 394, 1973 (Paper I)

\bibitem{Karami} Karami K., Bahari K., 2010, Sol. Phys., 263,
87

\bibitem{Karami} Karami K., Bahari K., 2012, ApJ, 757, 186


\bibitem{mort2009} Morton R.J., Erd\'{e}lyi R., 2009, A$\&$A, 502, 315

\bibitem{nak1999} Nakariakov V.M., Ofman L., DeLuca E.E., Roberts B., Davila J.M., 1999, Science, 285, 862

\bibitem{ofman2002} Ofman L., Aschwanden M.J., 2002, ApJ, 576, L153

\bibitem{Ofman1994} Ofman L., Davila J.M., Steinolfson R.S., 1994, ApJ, 421,
360

\bibitem{poed1989} Poedts S., Goossens M., Kerner W., 1989, Sol. Phys., 123, 83


\bibitem{Safari2006} Safari H., Nasiri S., Karami K., Sobouti Y., 2006, A\&A, 448, 375

\bibitem{Safari2007} Safari H., Nasiri S., Sobouti Y., 2007, A\&A, 470, 1111

\bibitem{sch2000} Schrijver C.J., Brown D.S., 2000, ApJ, 537, L69

\bibitem{tirry1996} Tirry W.J., Goossens M., 1996, ApJ, 471, 501

\bibitem{Van Doorsselaere} Van Doorsselaere T., Nakariakov
V.M., Verwichte E., 2007, A\&A, 473, 959

\bibitem{Verth} Verth G., Erd\'{e}lyi R., Jess D.B., 2008, ApJ, 687,
L45




\bibitem{ver2004} Verwichte E., Nakariakov V.M., Ofman L., Deluca E.E., 2004, Sol. Phys., 223, 77

\end{thebibliography}
\end{document}